**Localized crystallization of Ce:YIG thin films on Si using $CO_2$ laser annealing for integrated nonreciprocal photonic device applications**

*Xinran Ji, Junxian Wang, Tianchi Zhang, Xuan Zhao, Di Wu, Zixuan Wei, Yizhi Chen, Jialong Wang, Lei Bi**

* Corresponding author, Email: bilei@uestc.edu.cn

Xinran Ji, Junxian Wang, Tianchi Zhang, Xuan Zhao, Di Wu, Zixuan Wei, Yizhi Chen, Jialong Wang, and Lei Bi*
National Engineering Research Centre of Electromagnetic Radiation Control Materials, University of Electronic Science and Technology of China, Chengdu 610054, China
E-mail: bilei@uestc.edu.cn

Xinran Ji, Junxian Wang, Tianchi Zhang, Xuan Zhao, Di Wu, Zixuan Wei, Yizhi Chen, Jialong Wang, and Lei Bi*
State Key Laboratory of Electronic Thin-Films and Integrated Devices, University of Electronic Science and Technology of China, Chengdu 610054, China

**Abstract:** Laser annealing (LA) technique has emerged as an effective method for localized crystallization of magneto-optical (MO) garnet thin films on semiconductor substrates. However, no studies have explored the crystallization and magneto-optical (MO) properties of cerium-substituted yttrium iron garnet (Ce:YIG, $Ce_1Y_2Fe_5O_{12}$) thin films for integrated photonic device applications using LA technique. In this study, we provide a comprehensive investigation into the laser annealing of Ce:YIG films deposited on $SiO_2$ substrates and silicon nitride photonic waveguides for integrated nonreciprocal photonic device applications. Garnet phase was successfully observed in films grown on $SiO_2$ substrates, and SiN waveguides with laser annealing of sputtered Ce:YIG films on top of a laser annealed $Y_3Fe_5O_{12}$ seed layer. The magneto-optical (MO) properties of Ce:YIG films on oxidized Si substrates were found to be comparable to those prepared by rapid thermal annealing (RTA). A Mach-Zehnder Interferometer (MZI) type optical isolator based on Ce:YIG film on SiN was fabricated, exhibiting a saturation Faraday rotation of -2317.7 ± 94.8 deg/cm and propagation loss of 188.2 ± 16.5 dB/cm. Isolation ratio of 27.1 dB and insertion loss of 10.1 dB were achieved at 1552.7 nm wavelength.

## 1 Introduction

The rapid advancement of artificial intelligence (AI) and the Internet of Things (IoT) has led to extensive research on photonic integrated circuits for data communication and optical computing applications.[1-7] Optical isolators, circulators, and switches based on magneto-optical (MO) devices play a crucial role in silicon photonic systems.[8-16] For near-infrared optical communication wavelengths, magnetic iron garnet thin films have been integrated on silicon for nonreciprocal photonic device applications.[9-16] These films can be monolithically deposited on Si, followed by high-temperature rapid thermal annealing reaching over 800 °C for crystallization.[17-20] However, the high-temperature annealing process is not compatible with the backend-of-the-line (BEOL) process of silicon photonic circuits, which requires the substrate temperature to be below 400 °C. Therefore, a localized annealing process is desired for garnet thin film crystallization.

To overcome the challenges associated with high-temperature annealing, laser annealing has emerged as a recent effective alternative. This technique generates a controllable, localized temperature field in the laser-irradiated area, leaving surrounding regions unaffected. Laser annealing has been extensively studied and applied in various fields, including semiconductors, metals, thin-film technology, and novel optoelectronic materials.[21-26] In recent years, several studies have focused on investigating the crystallization and magneto-optical (MO) properties of garnet thin films using laser annealing.[27-31] For example, Sgibnev et al. employed low-energy, high-repetition-rate femtosecond laser pulses to crystallize Bi: YIG films, achieving a Faraday angle of 11 deg/μm at 515 nm.[29] Hibiki et al. utilized laser annealing in a vacuum chamber to fabricate Ce:YIG films. They successfully obtained crystallized Ce:YIG films with Faraday rotation angles of -0.25 deg/μm at 1064 nm and -0.04 deg/μm at 1550 nm.[30] In our previous study, we systematically explored the effects of laser annealing process parameters on the crystal structure and MO properties of YIG films.[31] YIG thin films were successfully crystallized by laser annealing on SOI waveguides and SiN micro-ring resonators, achieving comparable propagation loss to rapid thermal annealed (RTA) films.[31]

Despite these progresses, laser annealed rare-earth iron garnets still show large absorption loss and inferior magneto-optical (MO) effect, limiting their application in silicon photonic devices. In particular, laser annealed Ce:YIG films, which are important for magneto-photonic device applications in the 1550 nm wavelength range, still suffer from cracks, secondary phases, and large absorption loss. Integrated photonic devices based on laser annealed Ce:YIG had not been reported so far. In this study, we investigated Ce:YIG thin films fabricated via room-temperature magnetron sputtering and $CO_2$ laser annealing. By utilizing a laser annealed YIG seed layer, the garnet phase was stabilized, resulting in high phase purity, crack-free Ce:YIG garnet films with high Faraday rotation.[32,33] The crystal structure, magnetic properties, and magneto-optical (MO) characteristics of the Ce:YIG films were systematically analyzed under different laser powers and oxygen partial pressures. Moreover, we characterized the propagation loss and Faraday rotation of the laser-annealed Ce:YIG thin films on SiN waveguides. A Mach-Zehnder interferometer (MZI) type optical isolator was fabricated on SiN-based laser annealed Ce:YIG films, demonstrating an isolation ratio of 27.1 dB and insertion loss of 10.1 dB at 1552.7 nm wavelength.

## 2 Experimental Section

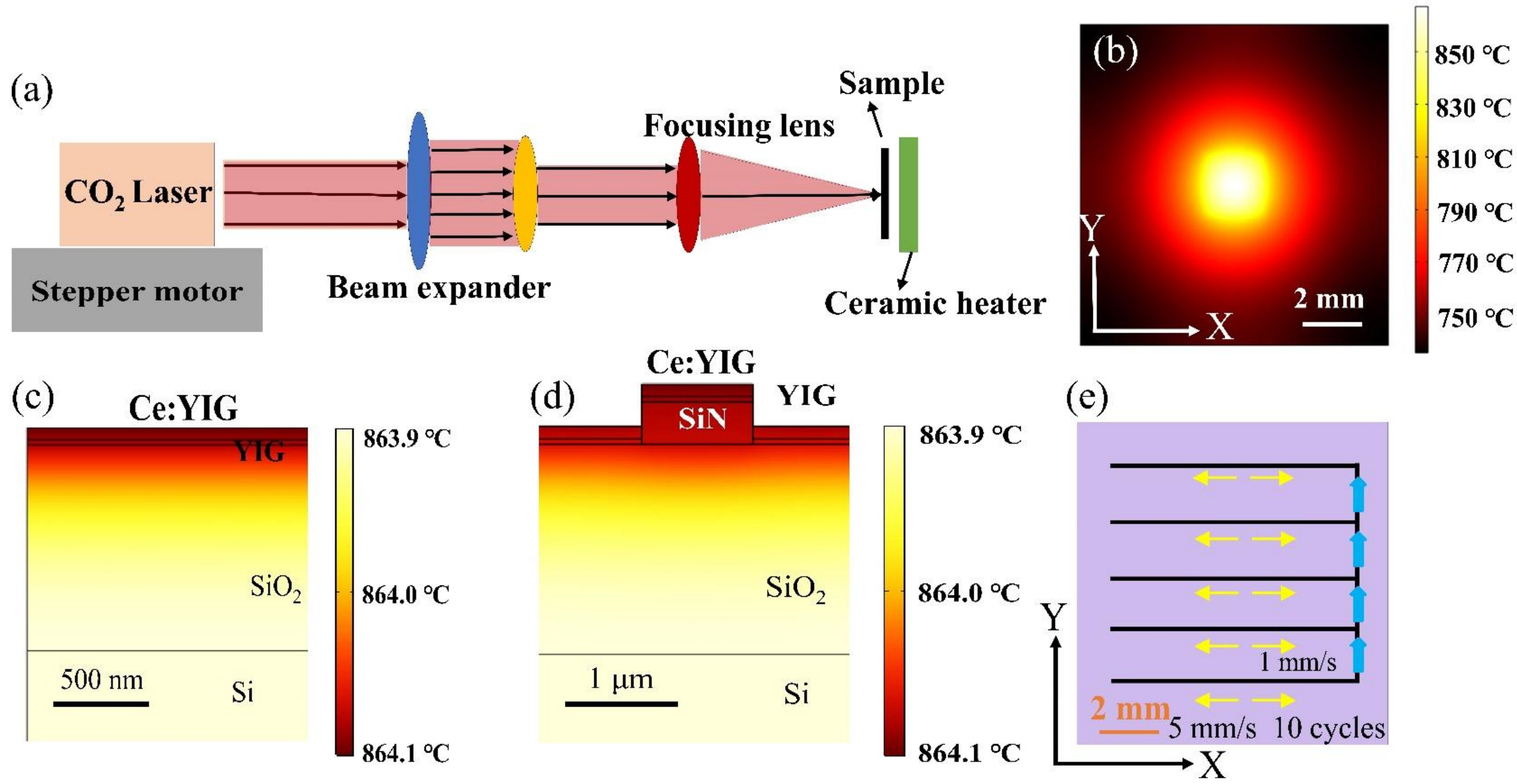

**Figure 1**. (a) Schematic diagram of the laser annealing apparatus. (b) Simulated temperature distribution of Ce: YIG thin films under $CO_2$ laser incidence with 1.8 mm × 1.8 mm spot size. Simulated temperature distribution at the sample cross-section when the incident laser wavelength is 10.6 μm for (c) the Ce:YIG/YIG/$SiO_2$ sample and (d) the Ce:YIG/YIG/SiN sample. (e) Laser scanning path on the chip.

YIG and Ce:YIG thin films were deposited onto Si substrates with 2 μm oxidized $SiO_2$ layer (RDMICRO) and silicon nitride (SiN) photonic waveguides using magnetron sputtering system (Leybold UNIVEX 400). The YIG ($Y_3Fe_5O_{12}$) and Ce:YIG ($Ce_1Y_2Fe_5O_{12}$) targets were synthesized via the standard solid-state reaction method using high-purity powders of $Y_2O_3$, $Fe_2O_3$ and $CeO_2$ (Sigma-Aldrich, 99.9%). The radio frequency (RF) power for sputtering was 90 W. The argon partial pressure and flow rate during sputtering were 0.5 Pa and 50 sccm, respectively. The substrate was kept at room temperature for both film deposition. The deposition rates of YIG and Ce: YIG were measured to be 0.55 nm/min and 0.5 nm/min, respectively. For convenience, material growth and laser annealing details are summarized in **Table 1**.

**Table 1**. Experimental parameters for YIG and Ce:YIG thin film deposited on silicon and SiN.

| Target | $Y_3Fe_5O_{12}$ | $Ce_1Y_2Fe_5O_{12}$ |
|---|---|---|
| Substrate temperature (°C) | Room Temperature | Room Temperature |
| Sputtering power (W) | 90 | 90 |
| Ar pressure (Torr) | 0.5 | 0.5 |
| Laser power ($W/mm^2$) | 2.78 | 2.31 - 4.17 |
| Oxygen partial pressure (Torr) | 2 | 0.5 - 2 |

**Figure 1(a)** illustrates the custom-built laser annealing apparatus, which consists of a $CO_2$ laser ((Synrad vi30+) wavelength of 10.6 μm and maximum power of 30 W), a beam expander, a beam homogenizer, and a ZnSe focusing lens (Thorlabs). This setup ensures that the laser spot is projected onto the surface of the sample in a square shape with an area of 1.8 mm × 1.8 mm. A ceramic heater was used to set the substrate temperature during laser annealing. The movement of the X and Y axes was controlled by a stepper motor (Feinixs). A 50 nm thick YIG seed layer was first deposited on $SiO_2$/Si substrates and SiN waveguides and fully crystallized using **the** $CO_2$ laser annealing technique, as described in our previous work.[31] 120 nm Ce:YIG films were then deposited on top of the YIG layer at room temperature by sputtering, followed by a subsequent laser annealing process. We first simulated the laser absorption process. The simulated temperature distribution of Ce:YIG thin films using a 1.8 mm × 1.8 mm laser spot (repeated scanning at a laser power of 3.24 $W/mm^2$) on a 1 cm × 1 cm chip is shown in **Figure 1(b)**. During the laser scanning, the temperature fluctuated around 830 °C, which ensured the crystallinity of the Ce:YIG film. **Figures 1(c)** and **(d)** show the temperature distribution at the sample cross-section for Ce:YIG/YIG/$SiO_2$ and Ce:YIG/YIG/SiN samples when the incident laser wavelength was 10.6 μm. Under the same incident conditions, the temperature difference of Ce:YIG films in steady state between samples with and without the SiN waveguide structure was small. Meanwhile, at a wavelength of 10.6 μm, the absorption coefficients of Si and Ce:YIG layers were much smaller than those of $SiO_2$. Therefore, for a given $SiO_2$ layer thickness, the Ce:YIG,YIG layer and Si structure on the sample surface had little effect on the temperature distribution. The simulation results also demonstrated that the presence of the YIG seed layer has no effect on the uniform temperature distribution of the upper Ce:YIG layer. During the laser annealing of Ce:YIG, several process parameters were varied, such as the laser power, oxygen partial pressure, substrate temperature, laser scanning cycles, and the speed of laser scans, to study their effects on the structure and properties of Ce:YIG. In this study, we kept the substrate temperature at 200 ºC. The laser scanning path was shown in **Figure 1(e)**. The scanning speed along the X-axis was 5 mm/s, and along the Y-axis was 1 mm/s. The number of X-axis scanning cycles was 10 cycles. We focused on the impact of laser power (2.31 W/mm$^2$ – 4.17 W/mm$^2$) and oxygen partial pressure (0.5 Torr – 2 Torr) on the Ce:YIG films.

The structure of the films was characterized using X-ray diffraction (XRD, Rigaku Co.) with a Cu-Kα radiation source. Microstructural images and crystallinity were analyzed using scanning electron microscopy coupled with electron backscatter diffraction (SEM-EBSD, JEOL JSM6460).[34] The root mean square (RMS) roughness was measured using an Atomic Force Microscope (AFM, Bruker). The electronic binding states of the films were assessed using X-ray photoelectron spectroscopy (XPS, Thermo Fisher Scientific K-Alpha), and the resulting data were analyzed using CasaXPS.[35-36] Magnetic hysteresis loops at room temperature were measured using a vibrating sample magnetometer (VSM, Lake Shore 7410). Faraday rotation hysteresis was measured using a custom-built system designed to characterize the Faraday effect. The SiN optical isolator devices were tested using a polarization-maintaining, fiber-butt-coupled system with TM-polarized incident light. A permanent magnet was employed to generate an external in-plane magnetic field of 1000 Oe, oriented perpendicular to the direction of light propagation, ensuring magnetic saturation of the garnet thin film in the MO/SiN waveguide. The transmission spectra were obtained following the procedure described in our previous work.[37] Backward transmittance measurements were performed by switching the input and output optical fibers connected to the polarization controller and optical power meter.

## 3 Results and discussion

### The influence of laser power

Ce:YIG films were laser annealed using different laser powers at an oxygen pressure of 1.5 Torr, after which their structure and magnetic properties were characterized. **Figure 2(a)** presents the X-ray diffraction patterns of the laser-annealed (LA) Ce:YIG films at various laser powers. Ce:YIG phases were observed at laser powers of 2.78 W/mm$^2$ and higher. No secondary phases were observable by XRD. The ionic radius of $Ce^{3+}$ (1.15 Å) was larger than that of $Y^{3+}$ (1.04 Å), which resulted in Ce:YIG diffraction peaks shifting to lower 2θ angles compared to YIG due to the incorporation of $Ce^{3+}$ ions into the crystal lattice.[13] **Figure 2(b)** illustrates the lattice constants of the Ce:YIG films after laser annealing at different laser powers. For an annealing power of 2.78 W/mm$^2$, the lattice constant of the film was relatively small, likely due to poor crystallization. The largest lattice constant was observed in Ce:YIG films annealed at 3.01 W/mm$^2$. Further increasing the laser power (≥3.01 W/mm$^2$), the lattice constant decreased. The decrease in lattice constant was attributed to the conversion of $Ce^{3+}$ to $Ce^{4+}$ ions at higher annealing temperatures. EBSD characterization shown in **Figure 2(c-d)** reveal high phase purity of Ce:YIG films. The highest crystallization ratio of the garnet phase reached 97.5±0.5% for Ce:YIG films annealed under 3.24 W/mm$^2$ (As shown in **Figure 2c**). However, when the laser power increased to 3.47 W/mm$^2$ or higher, the garnet phase purity decreased. This was due to the segregation of the impurity phase of $CeO_2$ and $Fe_2O_3$ under high laser power. **Figures 2(e)** displays the root mean square (RMS) roughness of the Ce:YIG films after laser annealing at different laser powers. The films annealed at 2.78-3.34 W/mm$^2$ exhibited low RMS roughness of 1.08 nm to 1.51 nm. When the laser power further increased, the RMS roughness of the Ce:YIG films increased to over 2 nm. This was attributed to the thermal mismatch between the YIG film ($10 \times 10^{-6}$/K) and the $SiO_2$ substrate ($0.47 \times 10^{-6}$/K), and the precipitation of $CeO_2$ and $Fe_2O_3$ phases.[19, 31]

To quantitatively study the Ce valence states, XPS characterization was performed.[38] **Figures 2(f)** shows the XPS Ce *3d* spectra of Ce:YIG films annealed at different laser powers. The Ce ions were primarily in the $Ce^{3+}$ and $Ce^{4+}$ states. In **Figure 2(f)**, for the sample annealed at 3.24 W/mm$^2$ laser power, eight peaks were observed. The Ce $3d_{3/2}$ and Ce $3d_{5/2}$ peaks appeared at 903.4 eV and 885.1 eV, respectively. The other six peaks, located at 882.1 eV, 898.3 eV, 889.4 eV, 900.7 eV, 907.2 eV and 916.6 eV corresponded to the $Ce^{4+}$ valence state, as shown in **Figure 2(f)**. The $Ce^{3+}$ concentration in the Ce:YIG thin films with different laser powers is shown in the inset of **Figure 2(f)**. As the laser power increased, the percentage of $Ce^{3+}$ at the film surface decreased from 50.1% to 11.54%, which is consistent with the XRD results, indicating that more $Ce^{3+}$ ions were oxidized to $Ce^{4+}$ at higher laser powers. According to thermodynamic principles, as the temperature rises, the change in free energy for the oxidation process becomes more favorable, promoting the oxidation of $Ce^{3+}$ to $Ce^{4+}$.[12, 39] Furthermore, the increased temperature accelerated the movement of oxygen molecules in Ce:YIG, which in turn increased the reaction rate and facilitated the oxidation process.

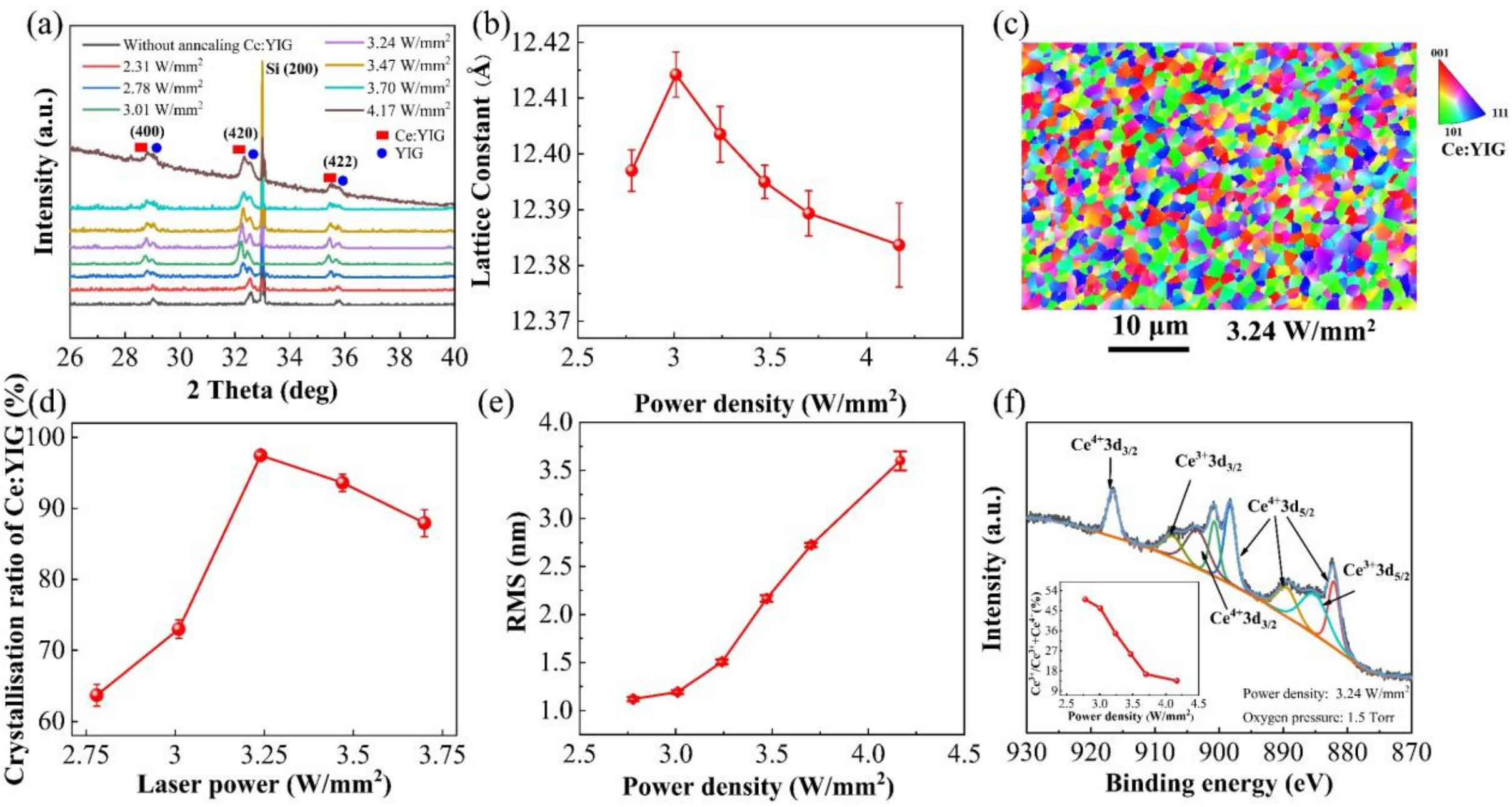

**Figure 2**. (a) X-ray diffraction patterns of Ce:YIG films annealed at different laser powers. The red squares and blue dots indicate Ce:YIG and YIG, respectively. (b) Lattice constants of annealed Ce:YIG film under different laser powers. (c) The EBSD (IPF map) of Ce:YIG film after 3.24 W/mm$^2$ power laser annealing. (d) The crystallization ratio of Ce:YIG films with different laser powers. (e) The root mean square value of roughness of Ce:YIG films after laser annealing at different laser powers. (f) High-resolution XPS spectra of Ce*3d* of Ce:YIG film annealed at 3.24 W/mm$^2$ laser power; The inset shows $Ce^{3+}$ concentration of Ce:YIG films as a function of laser power.

**Figures 3(a-d)** show the room-temperature Faraday rotation (FR) hysteresis loops and saturation FR of Ce:YIG thin films with different laser powers at wavelengths between 1200 nm and 1550 nm. The results indicated that increasing the laser power first caused an increase, then decrease in the FR. The highest FR values were -6573 ± 375 deg/cm at 1200 nm and -2544 ± 310 deg/cm at 1550 nm, achieved with a laser power of 3.24 W/mm$^2$. These values were close to the FR of RTA-Ce:YIG films on $SiO_2$ substrates annealed at 950 °C with $PO_2$ = 1.5 Torr for 100 seconds, which were -6316 ± 198 deg/cm at 1200 nm and -2416 ± 276 deg/cm at 1550 nm.[19, 40-41] The increase in FR with laser power was primarily attributed to the improved crystallinity of Ce:YIG. However, at higher laser powers, the increased $Ce^{4+}$ concentration enhanced optical absorption[40] and affected the charge-transfer transition from $Ce^{3+}$ (*4f*) to $Fe^{3+}$ (*3d*, tet.), leading to a reduction in the FR value.

**Figures 3(e)** displaya the room-temperature magnetic hysteresis loops of the Ce:YIG films at different laser powers. The in-plane (IP) and out-of-plane (OP) coercivity ($H_c$) values were relatively low, for instance, 4.8 Oe and 19.8 Oe at 3.24 W/mm$^2$, respectively. **Figure 3(f)** shows the saturation magnetization ($M_s$) and the corresponding out-of-plane saturation field ($H_s$) of the Ce:YIG films with laser powers ranging from 2.78 W/mm$^2$ to 3.70 W/mm$^2$. Compared to the in-plane magnetization, the out-of-plane magnetic field required to achieve saturation magnetization was approximately 2 kOe or higher, indicating that the material exhibited easy-plane magnetic anisotropy. The saturation magnetization initially increased and then decreased as the laser power rose. At the laser power of 3.24 W/mm$^2$, the film reached the highest saturation magnetization of 146.3 emu/cm³, attributed to the higher crystallinity of the film. However, at laser powers of 3.47 W/mm$^2$ and 3.70 W/mm$^2$, the $M_s$ values dropped to 110.3 emu/cm³ and 59.0 emu/cm³, respectively, which were lower than that of YIG. The changes in the net magnetic moment of Ce:YIG thin films were closely related to the interaction between the 4f orbital electrons of $Ce^{3+}$ and the *3d* orbital electrons of $Fe^{3+}$.[17, 41-42] Additionally, the magnetic moment of $Ce^{3+}$ ions is parallel to the minority $Fe^{3+}$ ions (16a). When the laser power was above 3.24 W/mm$^2$, $M_s$ decreased, which was attributable to the decrease in the crystallinity of Ce:YIG.

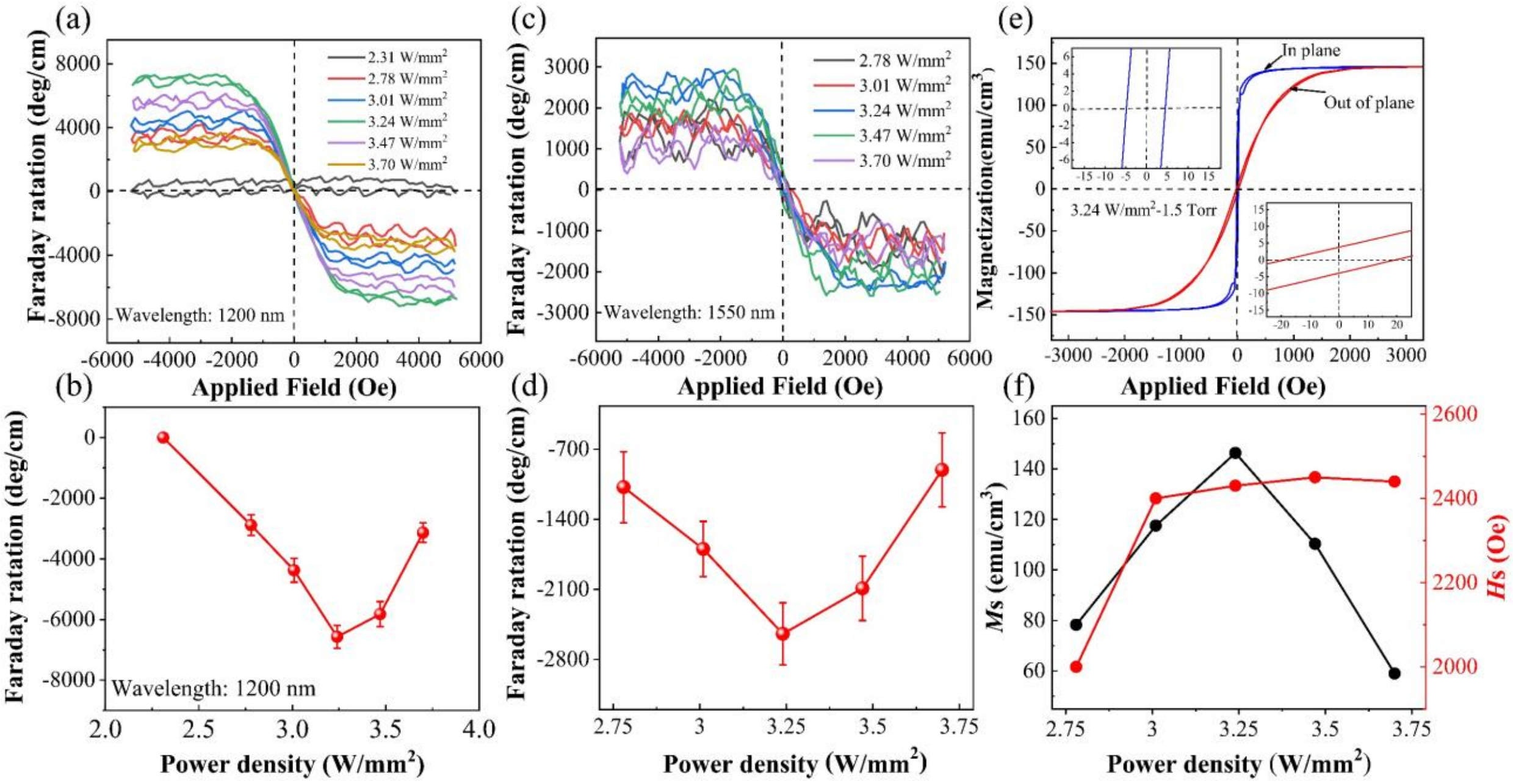

**Figure 3**. (a) FR hysteresis loops of Ce:YIG films annealed at different laser powers measured at the wavelength of 1200 nm. (b) Saturation FR of Ce:YIG films at 1200 nm. (c) FR hysteresis loops of Ce:YIG films annealed at different laser powers measured at the wavelength of 1550 nm. (d) Saturation FR of Ce:YIG films at 1550 nm. (e) The room temperature magnetic hysteresis of Ce:YIG films with laser annealing power of 3.24 W/mm$^2$. (f) The $M_s$ and out-of-plane saturation field $H_s$ of Ce:YIG films annealed with different laser powers.

### 3.2 The influence of oxygen partial pressure

Oxygen partial pressure is a crucial parameter that influences both the structural and magneto-optical (MO) properties of Ce:YIG films. To explore its impact, we fixed the laser power at 3.24 W/mm$^2$ for the laser annealing of Ce:YIG films under varying oxygen partial pressure conditions. **Figure 4(a)** shows the XRD patterns of the LA-Ce:YIG films at different oxygen partial pressures, revealing well-defined peaks that indicate the presence of a highly crystalline garnet phase in both the YIG and Ce:YIG layers. **Figure 4(b)** illustrates the variation in lattice constants of Ce:YIG thin films at different oxygen partial pressures under a laser power of 3.24 W/mm$^2$. The lattice constants of Ce:YIG films decreased as oxygen partial pressure increased, likely due to the reduction of oxygen vacancies and an increase in $Ce^{4+}$ concentration under oxygen-rich conditions.[43-45] **Figure 4(c)** shows the root mean square (RMS) roughness of the LA-Ce:YIG films annealed under different oxygen partial pressures. At lower oxygen partial pressures (0.5 Torr and 1 Torr), the RMS roughness remained largely unchanged. However, a significant increase in RMS roughness was observed at higher oxygen partial pressures, likely due to the formation of equilibrium phases, such as cerium oxide ($CeO_2$), at the grain boundaries.[44-46]

**Figures 4(d-e)** show the phase map, inverse pole figure (IPF) map, and crystallization ratio of Ce:YIG annealed under different oxygen partial pressures, obtained by EBSD characterization. The crystallization ratio of the LA-Ce:YIG films indicated that lower oxygen pressure was detrimental to the crystallization of Ce:YIG films. This was likely due to the insufficient oxygen concentration at lower pressures, which reduced the oxygen diffusion rate and increased the number of oxygen vacancies. As a result, the crystal growth and oxidation reactions were not effectively promoted, leading to incomplete crystallization and hindered crystal growth.[47-48]. **Figures 4(f)** presents the Ce (*3d*) XPS results of Ce:YIG films annealed under different oxygen partial pressures at 3.24 W/mm$^2$, revealing that at higher oxygen pressures, the redox reaction favored the oxidation process. Oxygen provided the necessary electrons to stabilize the $Ce^{4+}$ state, making the conversion of $Ce^{3+}$ to $Ce^{4+}$ more likely.[44-45]

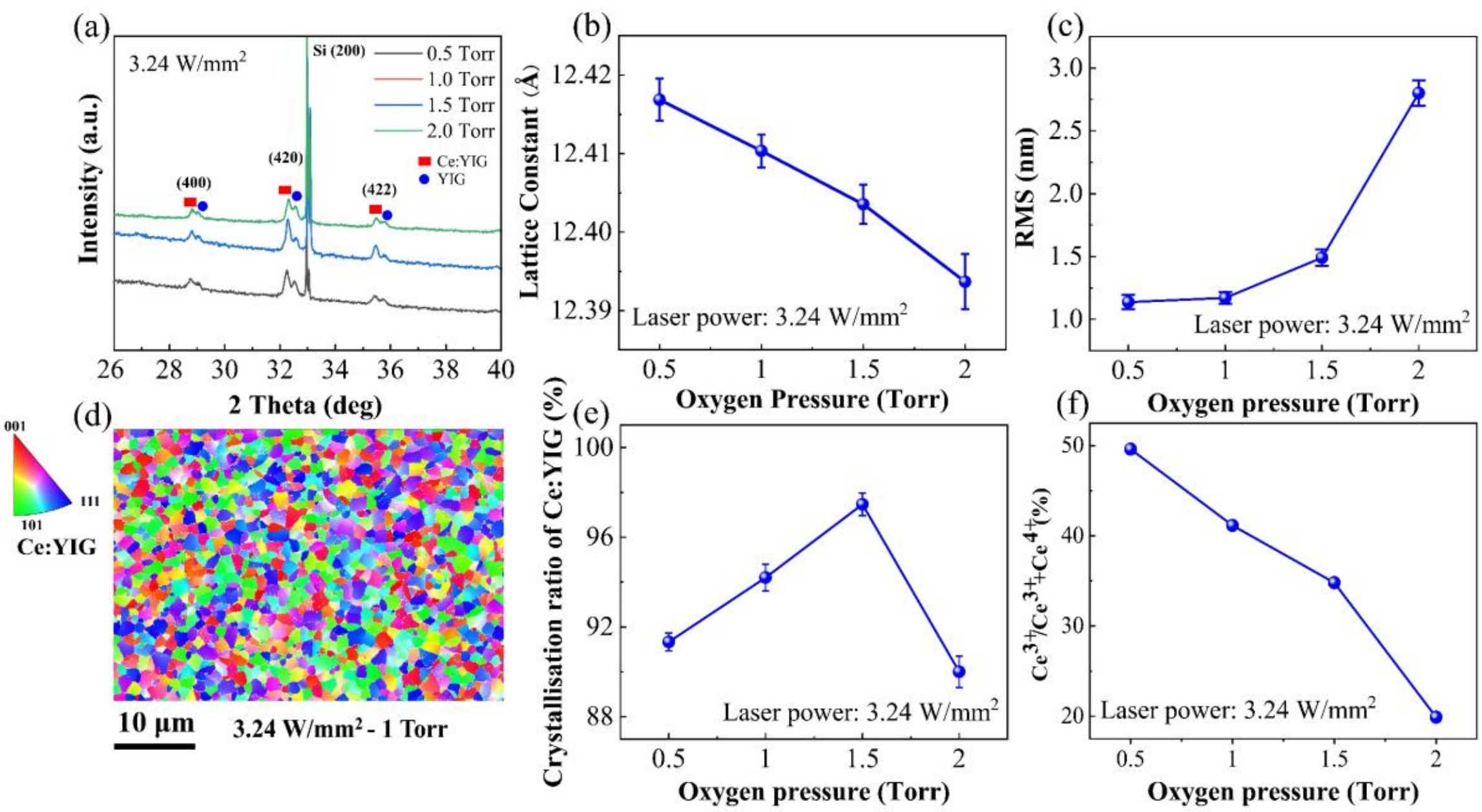

**Figure** 4. (a) X-ray diffraction patterns of Ce:YIG films annealed at different oxygen partial pressures. The red squares and blue dots indicate Ce:YIG and YIG, respectively. (b) Lattice constants of annealed Ce:YIG film under different oxygen partial pressures. (c) The RMS roughness of Ce:YIG films after laser annealing at different oxygen partial pressures. The inset showed an AFM image of YIG film after 3.24 W/mm$^2$ power laser annealing with 1.5 Torr oxygen partial pressure. (d) The EBSD after 3.24 W/mm$^2$ power laser annealing with 1 Torr oxygen partial pressure. (e) The crystallization ratio of Ce:YIG films with different oxygen partial pressures. (f) The $Ce^{3+}$ concentration of Ce:YIG films as a function of oxygen partial pressure.

Room-temperature Faraday rotation (FR) hysteresis loops for Ce:YIG thin films at wavelengths of 1200 nm and 1550 nm under different oxygen partial pressures are shown in **Figures 5(a)** and **5(c)**. The corresponding saturation FR values from these hysteresis loops are presented in **Figures 5(b)** and **5(d)**. The Faraday rotation of the films was strongly influenced by the annealing oxygen partial pressure. At very low oxygen pressures (0.5 Torr, 1 Torr), the Ce:YIG crystallization fraction decreased, while at excessively high oxygen pressure (2 Torr), the $Ce^{4+}$ content increased, both of which contribute to a reduction in the observed Faraday rotation value.[48] The oxygen partial pressure's effect on saturation magnetization ($M_s$) and corresponding out-of-plane saturation field ($H_s$) is highlighted in **Figure 5(e)**. The highest saturation magnetization value of $M_s$ = 146.3 emu/cm$^3$ was obtained at $PO_2$ = 1.5 Torr. This could be attributed to the incomplete crystallization of Ce:YIG at low oxygen pressures and the increased $Ce^{4+}$ content at high oxygen pressures, both of which negatively impact the magnetic

properties.[50]

The optical properties, including the refractive index and extinction coefficient in the near-infrared range, are crucial parameters for photonic devices.[19, 31, 40] These properties are also influenced by laser power and oxygen partial pressure. **Figures 5(f)** shows the refractive index and extinction coefficient of LA-Ce:YIG films under different laser powers and oxygen partial pressures. The refractive index of Ce:YIG films initially decreased and then increased with increasing laser power and oxygen pressure, respectively. Additionally, the extinction coefficients of all samples were relatively low following the laser annealing process. Under a 3.24 W/mm$^2$ laser power and 1.5 Torr oxygen partial pressure, the refractive index and extinction coefficient of the Ce:YIG layer were 2.21 at 1550 nm and 0 at 1550 nm, respectively. Note that the extinction ratio of Ce:YIG films could not be accurately measured by ellipsometry, indicating a relatively low loss of the fabricated film.

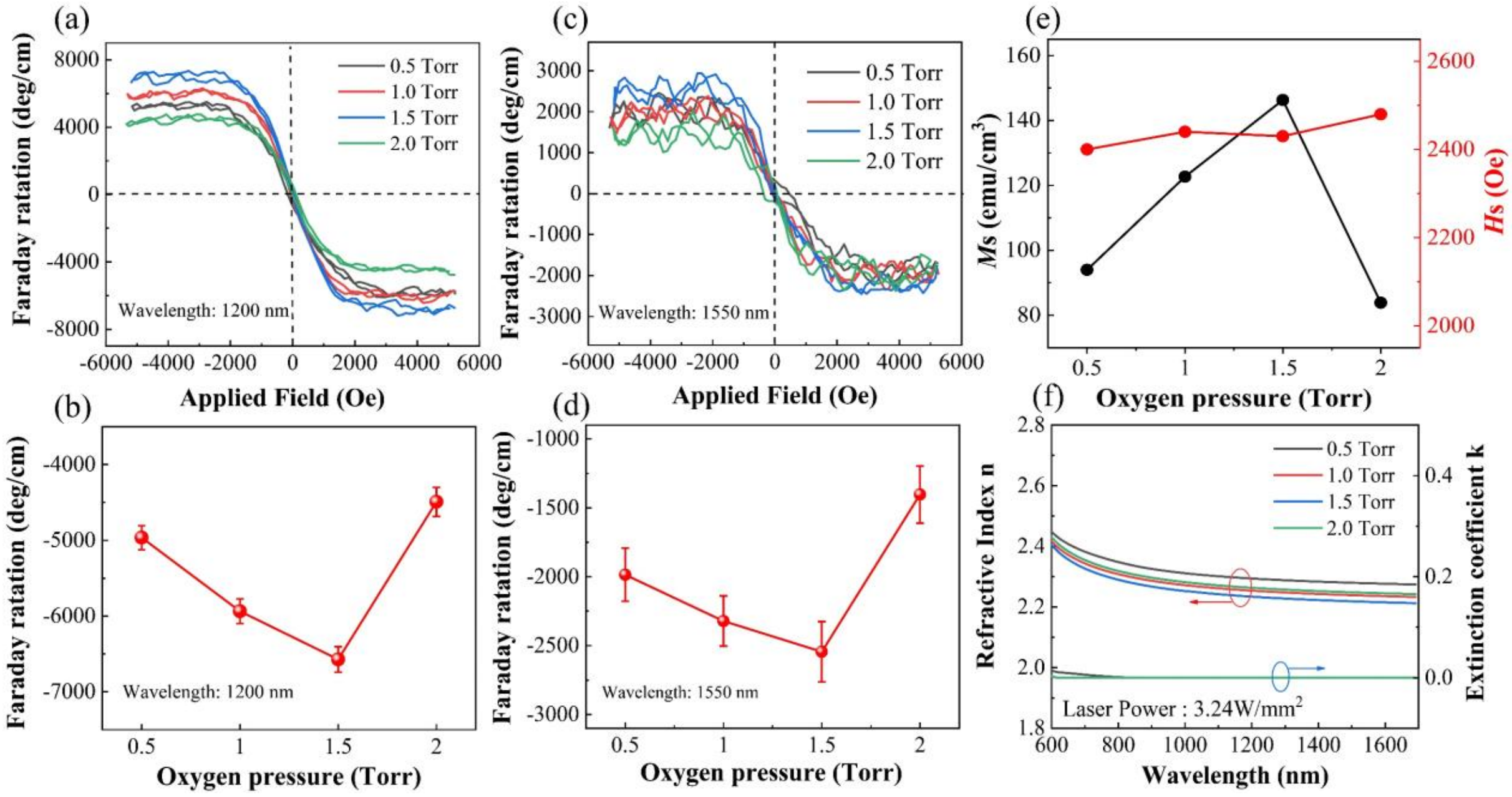

**Figure 5**. (a) FR hysteresis loops of Ce:YIG films annealed at different oxygen partial pressures measured at the wavelength of 1200 nm. (b) Saturation FR of Ce:YIG films at 1200 nm. (c) FR hysteresis loops of Ce:YIG films annealed at different oxygen partial pressures at the wavelength of 1550 nm. (d) Saturation FR of Ce:YIG films at 1550 nm. (e) The *M*s and out-of-plane saturation field *H*s of Ce:YIG films with different oxygen partial pressure. (f) Refractive index and Extinction coefficient spectra of Ce:YIG films characterized by spectroscopic ellipsometry with different oxygen partial pressure.

### 3.3 SiN integrated optical isolator based on laser annealed Ce:YIG

To achieve localized laser annealing, the thermal area with a temperature exceeding 400 °C

was reduced to 2.3 mm in diameter for subsequent film processing. The corresponding temperature distribution affected was monitored using an infrared CMOS camera, as shown in **Figure 6(a)**. The inset of **Figure 6(a)** shows the one-dimensional temperature profile (lateral position versus temperature) extracted from the infrared CMOS image, the 400 °C threshold, crystallization temperatures (>850 °C) of Ce:YIG films were clearly indicated. Based on the temperature profile, the diameter of the region exceeding 400 °C was estimated to be approximately 2.3 mm.

Next, we used the reduced thermal area to scan the Ce:YIG film (serpentine scan, 1 cycle, scan speed 50 μm/s), and the SEM image is shown in **Figure 6(b)**. It can be clearly observed that the scanning area (highlighted by the yellow area) exhibited well-defined crystallization, while the unexposed areas remained amorphous. From the SEM analysis, the overall crystallized area extended over several hundred micrometers due to the accumulation induced by the serpentine scanning path. Furthermore, analysis of individual laser scanning tracks indicated that the effective region reaching the crystallization temperature (>850 °C) during a single scan was approximately 58 μm in width. A comparison of SEM results and infrared thermal analysis results showed that the crystallization region was spatially confined to a narrow thermal zone. Only the central high-temperature zone (>850°C) played a role in the effective crystallization of garnet. These results indicated that the heat-affected zone during laser annealing scanning was crucial for the actual crystallization zone and the BEOL compatible process.

The present study successfully reduced the >400 °C thermal-affected region from approximately 20–30 mm under the previous 1.8 mm diameter laser spot condition to approximately 2.3 mm using a 200 μm diameter laser spot, corresponding to nearly an order-of-magnitude improvement in thermal localization. These results further suggested that the thermal region could potentially be confined to a scale approaching 200 μm through additional reduction of the laser spot size. However, further miniaturization was currently limited by the optical focusing capability of the existing experimental system.

In addition to reducing the laser spot size, several other approaches may further suppress the thermal-affected region during localized annealing. First, the use of visible-wavelength lasers had been reported as an effective strategy for improving thermal localization due to the shallower optical absorption depth and higher absorption efficiency in thin films.[27,51] Second, current-induced annealing might provide an alternative route for highly localized heating. Compared with optical heating, Joule heating generated by microfabricated electrodes could spatially confine the thermal region more effectively.[52] Third, thermal management of the substrate stage was also critical for minimizing heat diffusion. In semiconductor laser processing, vacuum-assisted thermal contact between the sample and the heat sink was a commonly used technique to enhance heat extraction efficiency and suppress lateral thermal spreading. Therefore, combining high-thermal-conductivity substrates with vacuum adsorption and localized laser annealing might further reduce the >400 °C thermal region. This strategy had been widely adopted in laser processing and semiconductor annealing systems to improve thermal confinement and process stability.[53-56]

We integrated small thermal area annealed Ce:YIG thin films on a SiN Mach-Zehnder Interferometer (MZI) type optical isolator devices. **Figure 6(c)** displays the device cross-sectional structure and the simulated electric field ($E_y$) distribution of the fundamental TM mode using COMSOL Multiphysics. Additional details on device design and fabrication can be found in our previous works.[13, 50]

Based on the simulation results, the confinement factors for Ce:YIG, YIG, and SiN films were calculated to be 11.37%, 6.17%, and 65.75%, respectively. The waveguide propagation loss can be calculated as:[31, 50]

$$\alpha_{WG} = \Gamma_{Ce:YIG} \cdot \alpha_{Ce:YIG} + \Gamma_{YIG} \cdot \alpha_{YIG} + \Gamma_{SiN} \cdot \alpha_{SiN} \qquad (1)$$

Where α represents the propagation loss for the Ce:YIG films ($\alpha_{Ce:YIG}$), YIG films ($\alpha_{YIG}$), and SiN waveguide ($\alpha_{SiN}$), respectively, while Γ denotes the confinement factors for these materials. Based on our previous reports, we neglected the loss of $SiO_2$ cladding layers, and the propagation loss of SiN was set to $\alpha_{SiN}$ = 0.5 dB/cm.[50] Optical microscope images of the MZI isolator on silicon nitride waveguides were shown in **Figure 6(d)**. EBSD results showed (As shown in the **Figure 6(e)**) that Ce:YIG films on silicon-based non-reciprocal devices obtained by small-spot laser scanning have excellent crystallinity (97.9%).

**Figure 6(f)** shows the transmission spectra of TM-mode SiN optical isolators with LA-Ce:YIG films using small-spot laser annealing. Clear one-way transmission was observed. The device achieved an isolation ratio of 27.1 dB with an on-chip loss of 10.1 dB at 1552.7 nm. The device exhibited an on-chip loss of 5.9 dB at 1565.1 nm. The on-chip loss included scattering loss of 1.7 dB due to the scattering at MO/SiN and SiN waveguide junctions, as well as SiN waveguide bending loss of 0.2 dB).[37, 50] The propagation loss of the laser-annealed YIG film was independently characterized before Ce:YIG deposition using a MZI device. Specifically, the insertion loss of the fabricated MZI device was experimentally measured, while the optical confinement factor in the YIG layer was obtained through optical mode simulations, enabling extraction of the YIG propagation loss. Based on the measured propagation loss, the non-reciprocal phase shifter (NRPS) length, the propagation loss of SiN ($\alpha_{SiN}$ = 0.5 dB/cm), the measured propagation loss of YIG ($\alpha_{YIG}$ = 75.9 ± 5.4 dB/cm), and the confinement factor, we calculate the Ce:YIG loss to be 188.2 ± 16.5 dB/cm. The measured NRPS value of the Ce:YIG film was found to be 5.78 rad/cm from the peak shift, corresponding to a Faraday rotation of approximately -2317.7 ± 94.8 deg/cm. This result was smaller compared to free-space measurement results, indicating different crystallinity of CeYIG films when crystallized on SiN. The Faraday rotation and optical propagation loss was inferior compared to RTA-Ce:YIG on SiN waveguides (propagation loss = 79 ± 5.0 dB/cm, Faraday rotation = -2500 deg/cm). During the annealing process of Ce:YIG films, issues such as atomic diffusion, non-uniform oxygen concentration distribution, and materials deviations from stoichiometric composition hindered the complete formation of the garnet structure. This led to the precipitation of thermodynamically more stable secondary phases, such as $CeO_2$ and $Fe_2O_3$, thereby degrading the performance of silicon-based nonreciprocal devices. In future work, we will focus on improving stoichiometric control of the material composition, introducing barrier layers, and exploring elemental substitution in Ce:YIG, with the aim of further enhancing the performance of nonreciprocal waveguide devices.

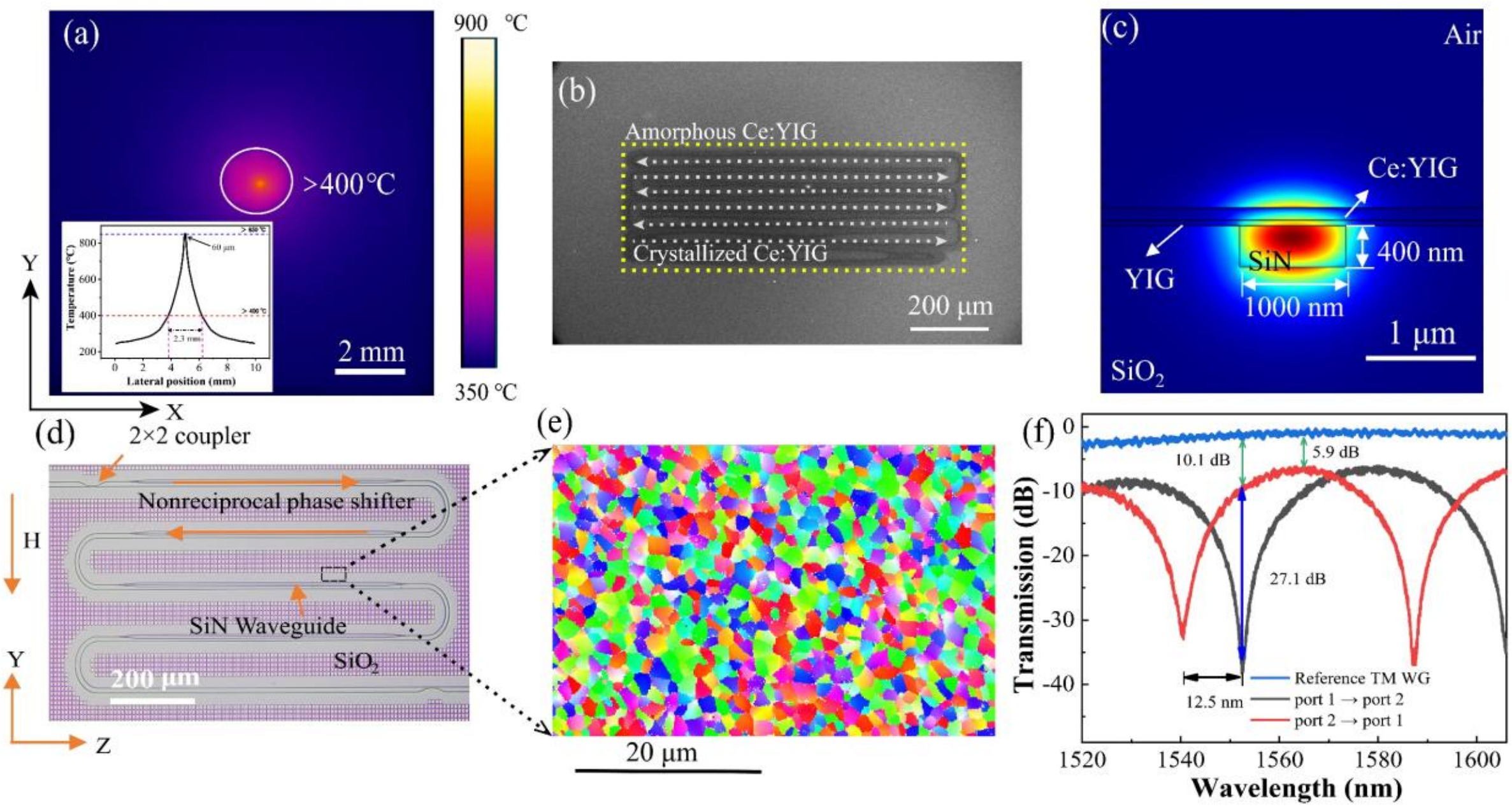

**Figure 6**. (a) Temperature distribution image captured by the infrared CMOS camera. (b) The SEM of crystallized Ce:YIG and amorphous Ce:YIG after focused $CO_2$ laser annealing. (c) Simulated Ey field distribution of the fundamental TM mode of the MO/SiN waveguide. (d) Optical microscope image of an MZI isolator on SiN waveguide. (e) The EBSD image of Ce:YIG showing different crystal orientation on the MZI isolator after focused $CO_2$ laser annealing. (f) The transmission spectrum of the MZI optical isolator after focused $CO_2$ laser annealing.

## 4. Conclusion

In summary, we successfully synthesized pure Ce:YIG films on silicon oxide substrates and SiN waveguide platforms using $CO_2$ laser annealing. By optimizing the laser power and oxygen partial pressure, the laser-annealed Ce:YIG films demonstrated excellent magnetic properties ($M_s$ = 146.3 emu/cm$^3$) and strong Faraday effect reaching -6573 ± 375 deg/cm at 1200 nm wavelength and -2544 ± 310 deg/cm at 1550 nm wavelength respectively. These properties were comparable to those of RTA-Ce:YIG samples. A SiN-based MZI-type optical isolator using LA-Ce:YIG was experimentally demonstrated, featuring a 27.1 dB isolation ratio and 10.1 dB insertion loss at 1552.7 nm wavelength. This work offers a backend compatible and monolithic integration strategy for integration of magneto-optical (MO) nonreciprocal photonic devices on silicon photonic circuits.

## Acknowledgements

The authors are grateful for the support by the National Natural Science Foundation of China (NSFC) (Grant Nos. 52450018, U22A20148, 52021001 and 52473292), Sichuan Provincial Science and Technology Department (Grant Nos. 2025ZNSFSC0040, 2024NSFSC0484).